\documentclass[11pt]{article}
\usepackage{graphicx,amsmath}

\renewcommand{\appendix}
        {
        \par
        \setcounter{section}{0}
        \setcounter{subsection}{0}
        \gdef\afterthesectionpunctdefault{:}
        \gdef\thesection{{Appendix \Alph{section}}}
        \renewcommand{\theequation}{\Alph{section}\arabic{equation}}
        \setcounter{equation}{0}
        }
\def\lsim{\hbox{\lower .8ex\hbox{$\, \buildrel < \over \sim\,$}}}
\def\gsim{\hbox{\lower .8ex\hbox{$\, \buildrel > \over \sim\,$}}}

\def\rp{\vec r_\perp}
\def\ep{{\mbox{\large e}}}

\newfont{\ensmathquatorze}{msbm10 scaled 1400}
\newfont{\ensmathonze}{msbm10 scaled 1100}
\newfont{\ensmathdix}{msbm10}
\newfont{\ensmathneuf}{msbm10 scaled 833}
\newfont{\ensmathhuit}{msbm10 scaled 694}
\newfam\ensmathfam                        
\textfont\ensmathfam=\ensmathonze        
\scriptfont\ensmathfam=\ensmathdix       
\scriptscriptfont\ensmathfam=\ensmathhuit

\begin{document}

\begin{center}

{\huge Breakdown of superfluidity of}\\

\

{\huge an atom laser past an obstacle}\\

\vspace{1.0cm}

{\Large Nicolas Pavloff}
\end{center}


\begin{center}
\noindent Laboratoire de Physique Th\'eorique
et Mod\`eles Statistiques\footnotemark,\\
\vspace{0.1 cm}
Universit\'e Paris Sud, b\^at. 100, F-91405 Orsay Cedex, France \\
\vspace{0.1 cm}
email: pavloff@ipno.in2p3.fr

\vspace{2 cm}
{\bf Abstract}
\end{center}
The 1D flow of a continuous beam of Bose-Einstein condensed atoms in the
presence of an obstacle is studied as a function of the beam velocity and of
the type of perturbing potential (representing the interaction of the obstacle
with the atoms of the beam). We identify the relevant regimes:
stationary/time-dependent and superfluid/dissipative; the absence of drag is
used as a criterion for superfluidity. There exists a critical velocity below
which the flow is superfluid. For attractive obstacles, we show that this
critical velocity can reach the value predicted by Landau's approach. For
penetrable obstacles, it is shown that superfluidity is recovered at large
beam velocity. Finally, enormous differences in drag occur when switching from
repulsive to attractive potential.

\vspace{3 cm}
\begin{math}
\footnotetext[1]{Unit\'e Mixte de Recherche de l'Universit\'e Paris XI et du
CNRS (UMR 8626).}
\end{math}

\noindent PACS numbers:
\vspace{0.5 cm}

\noindent 03.75.-b Matter waves\hfill\break
\noindent 05.60.Gg Quantum transport\hfill\break
\noindent 42.65.Tg Optical solitons; nonlinear guided waves\hfill\break
\noindent 67.40.Hf Hydrodynamics in specific geometries, flow
in narrow channels\hfill\break
\newpage

\section{Introduction}

	The rapid progresses in the technology of guiding cold atoms (using hollow
optical fibers \cite{fiber}, magnetic guides \cite{magnet} and microchips
\cite{chip}) opens up the prospect of similar studies of guided Bose-Einstein
condensed beams, i.e., of guided continuous atom lasers. Indeed, important
progress in this direction is presently being made, which uses the techniques
developed for guiding cold atoms: wave guides have been designed for a
Bose-Einstein condensate using a blue-detuned hollow laser beam \cite{Bon01};
Bose condensation has been obtained over a microchip \cite{Han01,Ott01} and a
Bose-Einstein wave-packet has been propagated in a microfabricated magnetic
wave-guide \cite{Lea02}. Also, a continuous beam of cold atoms has been loaded
into a magnetic guide, as a first step in order to perform evaporative cooling
and condensation in the guide \cite{Cren02}.

\

	In the present work we address the question of superfluidity of a continuous
(guided) atom laser, namely: what are the conditions for the flow to be
dissipationless ? A criterion for superfluidity has been proposed long ago by
Landau \cite{Kha00} which states that dissipation doesn't occur if the
velocity of the flow is lower than the critical value $v_{crit}=
\,\mbox{min}\,\{E(q)/q\}$, where $E(q)$ is the energy of an excitation with
momentum $q$.

	Many experiments have been done in liquid helium II trying to test Landau's
idea, and indeed one finds a critical velocity, but, in many instances it is  
much lower than Landau's expectation. As Feynman first suggested \cite{Fey57},
this is linked to vortex formation, i.e., to non-linear perturbation of the
fluid (and not to elementary excitations as implied by Landau \cite{Cli95}).
The experiments done at M.I.T. for Bose-Einstein condensates confirm this
view: in these systems there is also a critical velocity \cite{Ram99,Ono00},
it is lower that Landau's expectation, and it is also linked to vortex
formation \cite{Ino01}.

\

  In the following we are concerned with one-dimensional flows, which are
relevant to atom laser physics. We devise an alternative phrasing of Landau's
argument based on perturbation theory and identify its limit of validity. In
the generic, non-perturbative case, we confirm that, for obstacles represented
by {\it repulsive potentials}, dissipation begins at a velocity lower than the
one expected on the basis of Landau's argument and corresponds to emission of
solitons (which are the one-dimensional analogues of vortices). However, we
show that Landau's critical velocity is always reached for {\it attractive
potentials}. Furthermore, following a previous study \cite{Leb01}, above
Landau's critical velocity we identify a new (numerically stable) regime,
stationary {\it and} dissipative. In this regime, the drag exerted on an
obstacle can be computed with little numerical effort. Moreover, in this
regime, we show that at large velocity the drag exerted on a penetrable
obstacle goes to zero, i.e., superfluidity is recovered.

\

	The paper is organized as follows~: In Sec.~2 we set up the theoretical
framework and notations. The natural criterion for the breakdown of
superfluidity is the absence of drag, and in this section we show precisely
how the drag can be computed. In Sec.~3 we determine the different types of
flow and the corresponding drag for an obstacle represented by an external
potential (a weak potential in Sec.~3.1, a delta-peak in Sec.~3.2, a square
well in Sec.~3.3 and a Gaussian potential in Sec.~3.4). Finally we discuss our
results in Sec.~4, where we emphasize the important differences between
attractive and repulsive potentials in the non-linear regime.

\section{A criterion for superfluidity}\label{sec2}

	We work in a quasi one-dimensional regime, or more precisely we use an
adiabatic approximation where the condensate wave-function $\Psi(\vec{r},t)$
can be cast in the form \cite{Jac98,Leb01}
\begin{equation}\label{e1}
\Psi(\vec{r},t) = \psi(x,t) \, \phi(\rp; n) \; , \end{equation}
\noindent where $\psi(x,t)$ describes the motion along the axis of the laser.
$\phi$ is the equilibrium wave function (normalized to u\-ni\-ty) in the
transverse ($\rp$) direction; it depends parametrically on the longitudinal
density $n(x,t)=\int d^2r_\perp |\Psi|^2=|\psi(x,t)|^2$. The beam is confined
in the transverse direction by a trapping potential $V_\perp(\rp)$. The
adiabatic approximation assumes that the transverse scale of variation of the
profile is much smaller than the longitudinal one. The transverse degrees of
freedom are not completely frozen, but adapt to the smooth longitudinal
dynamics: this is the essence of the parametric dependence of $\phi$ on
$n(x,t)$. This represents a significant improvement to what is generally
defined as quasi one-dimensional approach and results in a non-typical
nonlinearity of the 1D reduction of the Gross-Pitaevskii equation below
(Eqs.~(\ref{e2},\ref{e3})).

\

	In the regime in which  Eq.~(\ref{e1}) holds \cite{attention}, the equation
governing the time evolution of $\psi(x,t)$ reads \cite{Jac98,Leb01} (we set
units such that $\hbar=m=1$):
\begin{equation}\label{e2}
-\frac{1}{2}\partial^2_{xx}\psi + 
\Big\{V_\parallel(x) + \epsilon[n(x,t)]\Big\}\,\psi =
 i\,\partial_t\psi \; .
\end{equation} 

  In (\ref{e2}), $V_\parallel(x)$ represents the effect of the obstacle. We
restrict ourselves to the case of a localized perturbation with
$\lim_{x\to\pm\infty} V(x)=0$. Such an obstacle can be realized by crossing
the trajectory of the atom laser with a detuned optical laser beam whose waist
is large compared with the perpendicular extension of the condensed beam. An
other possibility is to bend the trajectory of the guided atom laser; this
results in an attractive effective potential proportional to the square of the
curvature (see \cite{Leb01}).

	$\epsilon(n)$ is a non-linear term describing the mean-field interaction
inside the beam, and the way it is affected by the transverse confinement. For
a transverse confining harmonic potential of pulsation $\omega_\perp$ one has
(see \cite{Leb01}):
\begin{equation}\label{e3}
\begin{array}{lll}
\epsilon(n) =  2\,\omega_\perp\,na_{sc} &
\quad\mbox{in the low density regime} \quad & n a_{sc}\ll 1  \; , \\ 
\epsilon(n) =  2\,\omega_\perp\,\sqrt{na_{sc}} & 
\quad\mbox{in the high density regime} \quad & n a_{sc}\gg 1  \; ,
\end{array}
\end{equation}
\noindent where $a_{sc}$ denotes the s-wave scattering length of the two-body
inter-atomic potential (we consider only the case $a_{sc}>0$, i.e.
a repulsive effective interaction). In the following we use a formalism
allowing to treat both the high and the low density regime, since both will
be of interest in future guided atom laser experiments.

\
	
	We want to characterize the superfluidity of the flow past the obstacle. To
this end we compute the drag $F_d$ exerted by the atom laser on the obstacle:
a finite drag implies dissipation, whereas $F_d=0$ corresponds to a
superfluid flow. $F_d$ is defined as

\begin{equation}\label{drag}
F_d(t)=\int_{-\infty}^{+\infty}\!\!\!
dx \; n(x,t)\,\frac{d\,V_\parallel(x)}{dx} \; .
\end{equation}

	This definition is quite natural: the force exerted on the obstacle is the
mean value of the operator $dV_\parallel(x)/dx$ over the condensate
wave-function. It is rigorously justified by the analysis below in term of
stress tensor (Eq.~(\ref{e5})).

	For analytical determination of the drag, we use the following
procedure: the 1D version of the stress tensor of the fluid is \cite{stress}
\begin{equation}\label{e4}
T(x,t)=-\,\mbox{Im}\,(\psi^*\partial_t\psi)+
\frac{1}{2}|\partial_x\psi|^2-\varepsilon[n(x,t)]-V_\parallel(x)\,n(x,t)
\qquad\mbox{where}\qquad
\varepsilon(n)=\int_0^n\epsilon(\rho) \, d\rho\; ,
\end{equation}
\noindent and its impulsion density is
$J(x,t)=\,\mbox{Im}\,(\psi^*\partial_x\psi)$ (in our units, it is also the
current density). By the conservation equation
$\partial_t J+ \partial_x T +n\;\partial_x V_\parallel=0$, the total impulsion
of the beam $P(t)=\int_{-\infty}^{+\infty}dx\,J(x,t)$ is related to $F_d$ by:
\begin{equation}\label{e5}
\frac{d\,P}{dt} =
T(-\infty,t)-T(+\infty,t) - F_d(t)\; .
\end{equation}
	The physical content of (\ref{e5}) is clear: $dP/dt$ equals the total force
exerted over the beam. One part of this force ($-F_d$) is due to the
potential, the other one is the stress on the boundaries of the beam (at left
and right infinity). Hence Eq.~(\ref{e5}) confirms the heuristic guess
(\ref{drag}); besides, it allows to determine the drag in a simple
fashion in the stationary regime where $T$ and $P$ are time independent:
\begin{equation}\label{e5bis}
F_d=T(-\infty)-T(+\infty) \quad\mbox{in the stationary regime} \; .
\end{equation}
	Hence, in the following, we devote particular
attention to stationary solutions of (\ref{e2}). They are of the form
$\psi(x,t)=\exp\{-i\,\mu\,t\}\, A(x)\, \exp\{ iS(x)\}$, with $A$ and $S$ real
functions. The density is $n(x)=A^2(x)$, the velocity $v(x)=dS/dx$ and the
current $J(x)=n(x)v(x)$ is a constant that we note $J_\infty$. From
Eq.~(\ref{e2}), the amplitude $A(x)$ obeys a Schr\"odinger-like equation
\begin{equation}\label{e6}
- \frac{1}{2} \frac{d^2A}{dx^2}
+ \left\{ V_\parallel(x) + \epsilon[n(x)] + \frac{J_\infty^{2}}{2\, n^2(x)}
 \right\}\, A(x) = \mu \, A(x) \, . 
\end{equation}

	As discussed in \cite{Leb01}, the radiation condition requires that
solutions of (\ref{e6}) have no wake far down-stream: long range perturbations
of the beam only occur up-stream. Hence the boundary conditions have to be
imposed down-stream: because of nonlinearity one cannot disentangle an
incident and a reflected part in the perturbed up-stream flow. In all the
following we consider a beam going in the negative $x$ direction, with
down-stream boundary conditions $n(x\to-\infty)=n_\infty$ and
$v(x\to-\infty)=-v_\infty$ (with $v_\infty>0$). Then, the chemical potential
is $\mu=v_\infty^2/2+\epsilon(n_\infty)$. In the following we refer to
$v_\infty$ as the beam velocity and to $c_\infty= (\left.n_\infty \,
d\epsilon/dn\right|_{n_\infty})^{1/2}$ as the sound velocity (the proper
denomination should be ``sound velocity evaluated at constant density
$n_\infty$''). We also express the lengths in units of the relaxation length
$\xi= [2\,\epsilon(n_\infty)]^{-1/2}$.

	For stationary flows, the stress tensor (\ref{e4}) reads
\begin{equation}\label{e7}
T(x)=\frac{1}{2}\left(\frac{dA}{dx}\right)^2+W[n(x)]-V_\parallel(x)\,n(x)
\qquad\mbox{where}\qquad
W(n)=-\varepsilon(n)+\mu\,n+\frac{J_\infty^2}{2n}
\; .
\end{equation}
	In regions where the spatial variations of $V_\parallel(x)$ are negligible,
$T(x)$ is a constant (as easily seen from (\ref{e6})).

\section{Determination of the drag}

\subsection{Perturbative solution}

	Let us first evaluate the drag experienced by the obstacle when the effects
of the potential $V_\parallel(x)$ on the flow can be treated perturbatively.
In this case, by adiabatically branching the potential, one can always find a
stationary solution with $A(x)= \sqrt{n_\infty} + \delta A(x)$ having the
correct boundary condition (i.e., verifying the radiation condition).
Introducing the notation $\kappa =2\, |v_\infty^2-c_\infty^2|^{1/2}$ one finds
(see \cite{Leb01})~:
\begin{equation}\label{e8} 
\delta A (x) = \left\{ \begin{array}{lcc}
-\frac{\sqrt{n_\infty}}{\kappa} 
\int_{-\infty}^{+\infty} V_{\parallel}(y) \exp\{-\kappa
|x-y|\}\, dy & \mbox{when} & v_\infty < c_\infty  \; , \\ & & \\
\frac{2\, \sqrt{n_\infty}}{\kappa} \int_{-\infty}^{x} V_{\parallel}(y)
 \sin\{\kappa (x-y)\}\, dy &
\mbox{when} & v_\infty > c_\infty  \; . \end{array}\right.
\end{equation}
	The asymptotic evaluation of (\ref{e8}) when $x\to\pm\,\infty$ allows to
compute the drag through (\ref{e5bis}). From (\ref{e7}) and (\ref{e8}),
$T(-\infty)$ and $T(+\infty)$ differ only in the supersonic case where
\begin{equation}\label{s1}
T(-\infty)=W(n_\infty) \qquad\mbox{and}\qquad
T(x\to+\infty)=T(-\infty) +
\frac{1}{2}\,\left(\frac{d\,\delta A(x)}{dx}\right)^2+
\frac{\kappa^2}{2}\,\delta A^2(x) \; ,
\end{equation}
\noindent with (always in the super-sonic regime) 
\begin{equation}\label{s2}
\delta A(x)\underset{x\to+\infty}{\longrightarrow} 
\frac{2\sqrt{n_\infty}}{\kappa}\;
\mbox{Im}\;\left\{\ep^{i\kappa x}\,\hat{V}_\parallel(\kappa)\right\}
\; ,
\end{equation}
\noindent and $\hat{V}_\parallel (\kappa) = \int_{-\infty}^{+\infty} dx
\exp(-i\kappa x) V_\parallel(x)$ is the Fourier transform of $V_\parallel(x)$.
One thus obtains
\begin{equation}\label{e9}
F_d=0 \quad \mbox{when} \quad v_\infty < c_\infty  \; , \qquad\qquad
F_d=-2\,n_\infty |\hat{V}_\parallel(\kappa)|^2
\quad \mbox{when} \quad v_\infty > c_\infty \; .
\end{equation}

	The gross behavior characterized by (\ref{e9}) is general: at low velocity
the flow is superfluid, whereas at high velocity dissipation occurs. This
corresponds to Landau's criterion which determines a critical velocity below
which the flow is dissipationless: $v_{crit}= \,\mbox{min}\,\{E(q)/q\}$, where
$E(q)$ is the energy of an excitation with momentum $q$. For our system $E(q)$
is given by the Bogoliubov dispersion relation
$E(q)=q\,(c_\infty^2+q^2/4)^{1/2}$ (see, e.g., \cite{Dal99}) and the Landau
critical velocity is then $v_{crit} = c_\infty$. Hence the present
perturbative approach is identical to Landau's criterion since both give the
same value of velocity for the onset of dissipation and have the same physical
content: excitation of small, non-localized, perturbations is allowed only
above $v_{crit}$.

	However Landau's criterion, as well as the reasoning leading to
Eq.~(\ref{e9}) are, by essence, perturbative. We show below that, as discussed
in the introduction, non-linear effects alter these simple perturbative views.

\subsection{delta peak potential}

\begin{figure}[thb]
\begin{center}
\includegraphics*[width=7cm]{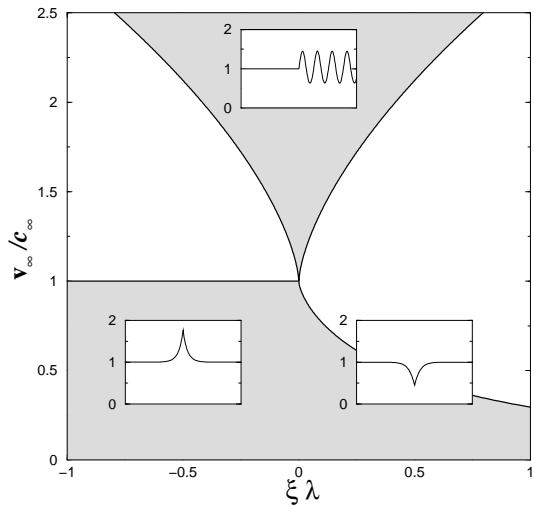}
\end{center}
\caption{\small\sl The shaded zone of the plane $(v_\infty,\lambda$) is the
domain of existence of stationary solutions occurring for a potential
$V_\parallel(x) = \lambda\,\delta(x)$. The axis are labelled in dimensionless
units. The insets represent density profiles $n(x)/n_\infty$ typical for the
different flows (the condensed beam is incident from the right). Each inset is
located at values of $v_\infty$ and $\lambda$ typical for the flow it
displays. The left (right) lower one is a superfluid flow across an attractive
(repulsive) potential. The upper one is a dissipative flow.}
\label{regimes_delta}
\end{figure}

	A first hint of the failure of the perturbative approach can be obtained if
studying the effect on the flow of a delta-potential
$V_\parallel(x)=\lambda\,\delta(x)$. In that case a stationary solution is
obtained by matching two free propagation modes of the laser (i.e., solutions
of (\ref{e6}) in the absence of potential) at $x=0$ with the condition
$A'(0^+)-A'(0^-) =2\lambda A(0)$. The upstream and downstream stress tensors
are constant, they are related by $T_{up}=T_{down}+\lambda
A(0)[A'(0^-)+A'(0^+)]$. Besides, at velocity $v_\infty>c_\infty$, the
radiation condition imposes that the down-stream ($x<0$) density is constant:
this gives $n(0)=n_\infty$ and $A'(0^-)=0$. Hence, for beam velocity
$v_\infty$ larger than $c_\infty$, in the stationary regime (\ref{e5bis})
yields:
\begin{equation}\label{e10}
F_d=-2\,n_\infty\,\lambda^2\; .
\end{equation}
	The same formula would have been obtained by using the perturbative result
(\ref{e9}). Hence it may seem that the perturbative approach is valid for all
range of delta potentials and velocities. However, one would expect
perturbation to fail for strong potentials and near $v_\infty=c_\infty$, since
it is meaningful only if $|\delta A(x)|\ll \sqrt{n_\infty}$, i.e., if
$|\lambda|\ll \kappa=2\, |v_\infty^2 - c_\infty^2|^{1/2}$ (see
Eq.~(\ref{e8})). Indeed, whereas a naive perturbative approach (leading to
(\ref{e8})) predicts that a stationary solution exists for all values of
$v_\infty$, an exact solution in presence of a delta-potential can be found
only in some precise range of velocity and potentials. This has been studied
in Ref.~\cite{Leb01} and is illustrated on Figure~\ref{regimes_delta}. Roughly
speaking, stationary regimes only exist for low and high values of $v_\infty$
(in the shaded zone of Fig.~\ref{regimes_delta}). In the low velocity case,
the stationary profile is a trough (or a bump, depending on the sign of the
potential) localized on the perturbation \cite{dis}, such flows are
dissipationless. At high velocity the stationary profile has a (non-linear)
wake extending to infinity in the upstream direction and this corresponds to
dissipation (by Eq.~(\ref{e5bis})). In between, the flow is time-dependent and
there is no reason why, in this case, the drag should be given by (\ref{e10}).

	We note here an important dissymmetry between repulsive ($\lambda>0$) and
attractive ($\lambda<0$) potentials: the domain of superfluid flow extends up
to $v_\infty=c_\infty$ for $\lambda<0$, i.e., the critical velocity has the
same value as the one predicted by Landau's approach. On the other hand, for
$\lambda>0$ the critical velocity is potential-dependent and decreases
significantly. This feature is very general (it occurs for all the other
potentials we have studied), we will comment on it more thoroughly in
Sec.~3.4 and in the Conclusion.

\subsection{Repulsive square well}

	An other case with analytical stationary solutions is the repulsive square
well: $V_\parallel(x)$ is zero, except for $0<x<\sigma$ where it takes the
constant and positive value $V_0$. In this case (\ref{drag}) yields:
\begin{equation}\label{e11}
F_d(t)=V_0\Big[ n(0,t) - n(\sigma,t) \Big] \; .
\end{equation}
		In the stationary subsonic regime, one can show that $n(\sigma)=n(0)$
(see \cite{Leb01}) and the drag is zero~: this characterizes a superfluid flow.
For $v_\infty>c_\infty$ instead, (\ref{e11}) yields a finite drag. In this
case, when the flow is stationary, $n(0)=n_\infty$ and $n(\sigma)$ in
(\ref{e11}) can be computed by quadrature, as a solution of
\begin{equation}\label{e12}
\sqrt{2}\,\left|\sigma-L_0\,\mbox{nint}\left(\frac{\sigma}{L_0}\right)\right|
 = \int_{\sqrt{n_\infty}}^{\sqrt{n(\sigma)}}
\frac{dA}{[T_0-W(n)+V_0\,n]^{1/2}}\; ,
\end{equation}
\noindent where nint denotes the nearest integer, $L_0$
is the period of density oscillations inside the well (which is again
expressible as an integral) and $T_0=W(n_\infty)-V_0\,n_\infty$ is the
constant value assumed by $T(x)$ inside the well. Of course, (\ref{e12}) is
only valid for a hypersonic stationary solution. In the regime where no
stationary solution exists, the drag is time-dependent and should be computed
numerically (as done in Section \ref{gp} below).

\

	Figure~\ref{force_carre} displays the evolution of the drag as a function of
the beam velocity $v_\infty$. The drag has been computed only in the
stationary regimes. In the subsonic stationary regime it is exactly zero
\cite{dis}. In the super-sonic stationary regime the exact result (\ref{e11})
(solid line) has been computed in two manners: through the numerical solution
of (\ref{e12}), and also by numerical integration of (\ref{e6}). Both methods
agree within 4 digits. The exact expression is compared in
Figure~\ref{force_carre} with the perturbative result (\ref{e9}) (where the
Fourier transform of the potential gives here $|\hat{V}_\parallel(\kappa)|^2 =
(2\,V_0/\kappa)^2\,\sin^2(\kappa\sigma/2)$). For large velocities, the
perturbative approach becomes more and more accurate. This is expected from
(\ref{e8}) since, when $\kappa\,\sigma\gg 1$ (i.e., at large velocities) the
perturbative solution is accurate if $V_0\ll\kappa^2$.

\begin{figure}[thb]
\begin{center}
\includegraphics*[width=8cm]{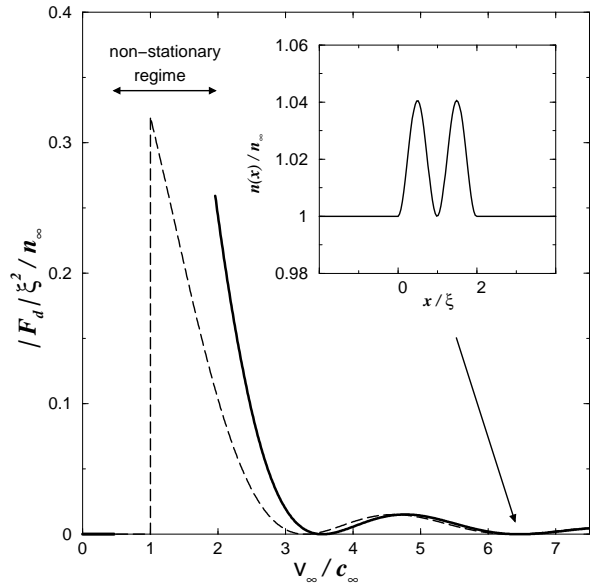}
\end{center}
\caption{\small\sl Drag exerted by a high density beam on a repulsive square
well ($V_0\,\xi^2=0.2$, $\sigma=2\,\xi$), as a function of the beam velocity
($F_d$ and $v_\infty$ are expressed in dimensionless units). The dashed line
is the perturbative result (\ref{e9}). The solid line is the exact drag
(\ref{e11}). It is evaluated here only in the stationary regime. The inset
represents the beam density in the supersonic regime, at a value of velocity
(indicated by the arrow) where the drag is exactly zero.}
\label{force_carre}
\end{figure}

	One can remark in Figure \ref{force_carre} that, even in the supersonic
regime, the drag happens to be exactly zero for some specific values of
$v_\infty$ \cite{Law00}. As illustrated in the inset, this occurs when the
period of the density oscillations inside the well is an exact divisor of its
width $\sigma$. In this case, the density is unperturbed outside the well. The
modification of density inside the well is minute in the case represented in
the Figure ($4\,\%$). However, for stronger potentials it can be quite
substantial: for $\sigma=2\,\xi$ and $V_0\,\xi^2=5$, the maximum density
inside the well reaches $2\,n_\infty$ while the density remains unperturbed
(i.e., equal to $n_\infty$) outside the well.

\subsection{Gaussian potential}\label{gp}

	The generality of the above deductions, based on the study of model
potentials (a delta peak and a square well) can be tested numerically on
more realistic potentials. We now consider the case $V_\parallel(x)=
V_0\exp\{-x^2/\sigma^2\}$ (with $V_0>0$). 

\begin{figure}[thb]
\begin{center}
\includegraphics*[width=8cm]{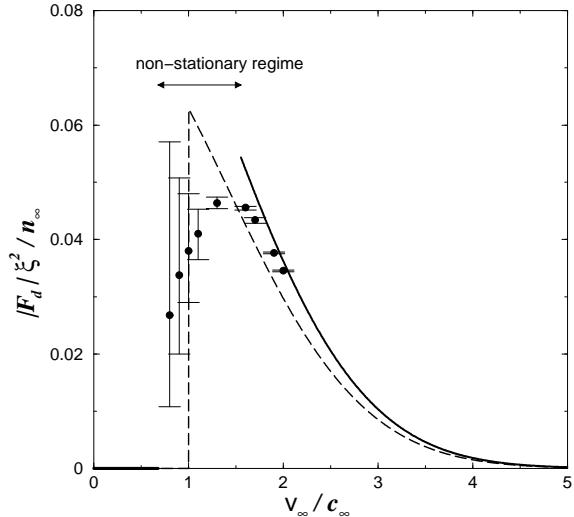}
\end{center}
\caption{\small\sl Drag exerted by a low density beam on a Gaussian potential
($V_0\,\xi^2=0.2$, $\sigma=0.5\,\xi$), as a function of the beam velocity.
The conventions are the same as in Fig.~\ref{force_carre}. The dashed line
is the perturbative result (\ref{e9}). The solid line is the exact drag
evaluated in the stationary regime. The circles correspond to the
drag evaluated in the time-dependent regime. The error bars correspond to
the extremal values of the time dependent $F_d(t)$ (see the text).}
\label{force3}
\end{figure}

	As in the previous cases, stationary solutions exist only if the beam
velocity $v_\infty$ is not too close to the sound velocity $c_\infty$. In the
subsonic stationary regime the density is perturbed only in vicinity of the
potential and the flow is superfluid (by Eq.~(\ref{e5bis})). In the supersonic
stationary regime, the density oscillations extend upstream to infinity, and
this corresponds to dissipation. At velocities where a stationary regime is
possible, we have determined the drag indifferently using (\ref{drag}) or
(\ref{e5bis}) (after having solved (\ref{e6}) numerically), whereas in the
non-stationary case we used (\ref{drag}) after having solved the time
dependent equation (\ref{e2}). The results are presented in Fig.~\ref{force3}.

	The behavior of the drag in the stationary regime confirms what is expected
from the results of the previous sections. In particular, the critical velocity
for the onset of dissipation is lower than $c_\infty$. The reason for this is
that, in the region of the repulsive obstacle, the density decreases and
conservation of flux requires that the local fluid velocity increases. As a
result, it may happen that Landau's criterion is verified (although
$v_\infty<c_\infty$) because the {\it local} fluid velocity reaches the sound
velocity. For slowly varying potentials (i.e., in the regime $\sigma\gg\xi$)
this argument was put on a firm mathematical basis by Hakim \cite{Hak97}.

	From this, one can infer that for {\it attractive potentials} instead, the
critical velocity should be equal to $c_\infty$ since for such potentials the
density increases in the region of the obstacle and the local velocity
decreases accordingly \cite{Fed01}. This has already been
observed in the case of a delta-peak potential (see Fig.~\ref{regimes_delta})
and also for an attractive square well (see Ref.~\cite{Leb01}). We have
performed numerical checks showing that the same occurs for an attractive
Gaussian potential. We will return in the final section to the difference
between attractive and repulsive potentials.

\begin{figure}[thb]
\begin{center}
\includegraphics*[width=8cm]{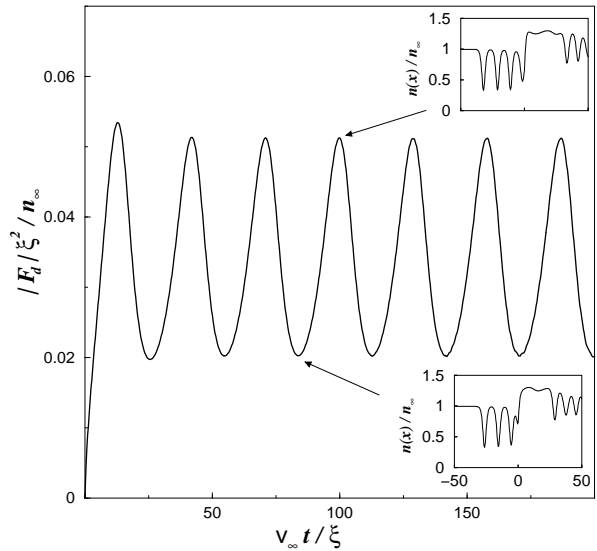}
\end{center}
\caption{\small\sl Time evolution of the drag exerted by a low density beam (of
velocity $v_\infty = 0.9\,c_\infty$) on a Gaussian potential ($V_0\,\xi^2 =
0.2$, $\sigma=0.5\,\xi$). $F_d$ and $t$ are expressed in dimensionless
units. The upper (lower) inset represents a density profile observed when
the drag is maximum (minimum).}
\label{force4}
\end{figure}

	Let us now come to the discussion of the time dependent data (the circles
of Fig.~\ref{force3}). They are drawn with error bars: this does not
correspond to a numerical uncertainty, but reflects the fact that the drag
depends on time in the non-stationary regime. The ``error bars'' correspond to
the extremal values of the time dependent function $F_d(t)$ (see
Fig.~\ref{force4}). The initial condition $\psi(x,t=0)$ was taken as the
stationary superfluid solution in the presence of the well for the sub-sonic
case $v^{init}_\infty = 0.5\,c_\infty$ to which a Galilean boost $\exp\{i\,
(v_\infty^{init}-v_\infty) \,x\}$ was applied at $t=0$ in order to reach the
desired value of velocity. Accordingly, in the computation, the drag $F_d(t)$
starts from 0 and after a set-up time reaches a regime where it oscillates
around a mean value (see Fig.~\ref{force4}) which is represented by the
circles in Fig.~\ref{force3}. The oscillations of $F_d(t)$ around its mean
value are of interest because they reflect the cause of time-dependence of the
flow: the numerics indicate that whereas the up-stream flow reaches a
quasi-stationary pattern, the down-stream density is perturbed by solitons,
periodically emitted from the obstacle, that propagate in the same direction
as the flow with a smaller velocity (such a behavior has already been observed
by Hakim \cite{Hak97}). As illustrated in Figure~\ref{force4}, the drag
decreases when a soliton has just been emitted (this was already noted in
two-dimensional flows by Winiecki {\it et al.} \cite{Win99}, where vortex
pairs are emitted from a moving obstacle).

\begin{figure}[thb]
\begin{center}
\includegraphics*[width=8cm]{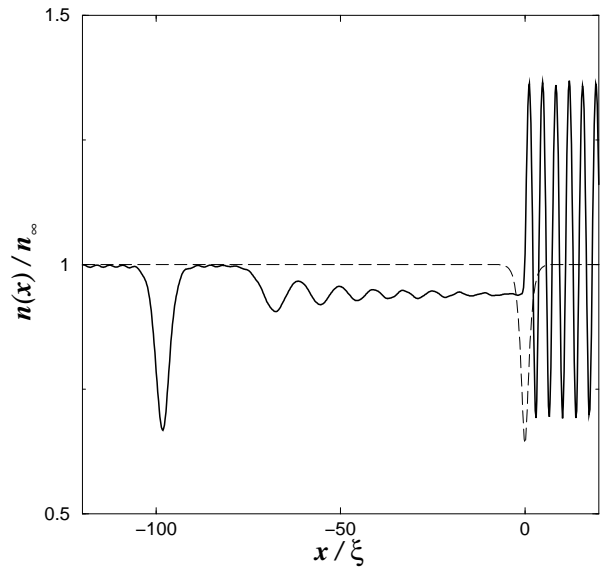}
\end{center}
\caption{\small\sl Density profile in the case $v_\infty=1.6\,c_\infty$, in
the `presence of a Gaussian potential ($V_0\,\xi^2 = 0.2$, $\sigma=0.5\,\xi$).
The plot represents the profile (solid line) after evolution of the initial
$n(x,t=0)$ (dashed line) during a time $t$ with $v_\infty\,t/\xi=203.6$. The
density trough initially located at $x=0$ has moved to the left and is now at
$x\simeq - 100\,\xi$. It will asymptotically move to left infinity. It is
followed by a density slightly depressed compared to the initial value
$n_\infty$. This depressed density is the asymptotic down-stream density.}
\label{asymp}
\end{figure}

	In the super-sonic regime, it is also important to discuss the discrepancy
between the results for the drag computed for stationary flows and for time
dependent ones (solid line and circles, respectively, in Fig.~\ref{force3}).
This discrepancy by no means implies that the stationary profile is unstable
or that the asymptotic time-dependent flow is not stationary. On the contrary,
numerics indicate that time-dependent flows reach an asymptotic
stationary regime for velocities at which such a regime exists. Moreover the
asymptotic density profile is of the expected type (flat down-stream and
oscillating up-stream). The point is that our specific initial condition
$\psi(x,0)$ does not asymptotically lead (when $t\to\infty$) exactly to
down-stream density and velocity which have the same value $n_\infty$ and
$v_\infty$ as the initial flow. This is illustrated in Figure~\ref{asymp}. On
the Figure one sees that the asymptotic down-stream density is about $6\,\%$
lower than $n_\infty$ and a simple numerical check shows as well that the
asymptotic velocity differs from $v_\infty$ (is is roughly $3\,\%$ higher).
This artifact becomes less and less important for increasing velocities: since
the perturbative approach is more and more accurate, it is clear that
down-stream modifications of the solution become minor and that the asymptotic
solution is the expected one. As a result, when $v_\infty$ increases, the
black circles get closer to the solid line in Figure~\ref{force3}.

\section{Discussion and conclusion}

	The above presented results illustrate the well known fact that non-linear
effects alter the simple perturbative (Landau's) approach for the
determination of the critical velocity at which dissipation occurs (we saw,
however, that the perturbative approach could reach a regime of accuracy at
large velocity). For repulsive potentials the critical beam velocity is
smaller than the velocity of sound $c_\infty$ (which is here the critical
velocity from Landau's criterion) because, in the region of the potential, the
local fluid velocity can reach values higher than the sound velocity. The
onset of dissipation corresponds to non-stationary flows with a wake
asymptotically extending upstream to infinity, and down-stream periodic
emission of solitons. In fact, another way of explaining the lower stability
of the dissipationless flow in the presence of a repulsive potential is to
remark that, for sub-sonic flow over such an obstacle, the density decreases
in the region of the potential, allowing easier nucleation of solitons.

	On the other hand, we have shown that, for attractive potentials, stationary
dissipationless solutions exist up to $v_\infty=c_\infty$: Bose-Einstein
condensates appear to be excellent supports for reaching Landau critical
velocity, more appropriate than superfluid helium because, in atomic vapors,
it is simpler to construct obstacles described by purely attractive potentials.

\

	We also showed, using analytical an numerical examples, that stationary
dissipative profiles exist in hyper-sonic beams, provided the beam velocity is
large enough. From the numerical study, these solutions seem stable, and
moreover time-dependent flows tend asymptotically to such solutions (when they
exist). It is interesting to note that dissipation is drastically reduced at
very high velocity: i.e., superfluidity is recovered. Such an effect should
also exist in higher dimensions for penetrable potentials. It can be
understood perturbatively: at high velocity, when the perturbative approach
becomes valid, the relevant wave-vector (denoted $\kappa$ in Sec.~3.1) is
large and Bogoliubov dispersion relation becomes exactly quadratic. Hence, in
this regime one has a matter-wave described by the linear Schr\"odinger
equation. The drag in this regime can be shown to be proportional to the
reflection coefficient which, as well known, decreases at high energy (in
any dimension).

\

 Note that, for the sake of clarity, we have always illustrated our
conclusions using rather weak perturbing potentials. The reasons for this are
twofold: (1) for stronger potentials the super-sonic stationary regime that we
wanted to emphasize is rejected to higher velocity; (2) the numerical effort
necessary in the study of the non-stationary regime is decreased for weak
potentials since the domain of time-dependence is reduced.

\begin{figure}[thb]
\begin{center}
\includegraphics*[width=8cm]{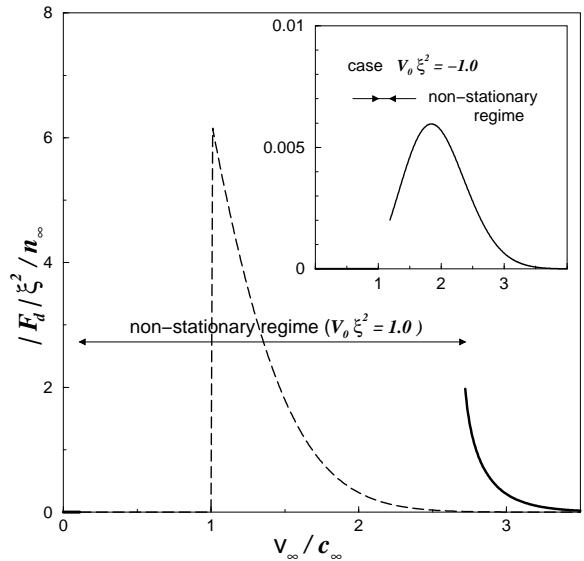}
\end{center}
\caption{\small\sl Drag exerted by a low density beam on a Gaussian potential
($V_0\,\xi^2=\pm 1.0$, $\sigma/\xi=1.0$), as a function of the beam velocity.
The conventions are the same as in Fig.~\ref{force_carre}. The main figure
displays the drag for a repulsive potential (solid line) together with the
perturbative result (\ref{e9}) (dashed line) which is not affected by the sign
of the potential. The inset displays a blow-up of the main figure allowing to
see the (very small) drag for an attractive potential.}
\label{att_rep}
\end{figure}

	Although for stronger potentials the qualitative results remain the same,
very interesting new quantitative phenomena occur. In particular, enormous
differences in drag can occur when switching from repulsive to attractive
potentials (provided the potential is strong enough). This is illustrated in
Fig.~\ref{att_rep}.

	One sees in the Figure that the domain of existence of stationary solutions
is markedly different between attractive and repulsive potentials and that the
values of the drag differ by more than two orders of magnitude ! The physical
explanation for this phenomenon is subtle. At high velocity, contrarily to
intuition, the density decreases (increases) in an attractive (repulsive)
potential \cite{trou}. Hence, an attractive potential creates a density
trough which, being the super-sonic analog of a gray soliton, does not create
large perturbations in its wake. This results in a small drag. This feature is
clearly missed by the perturbative approach which gives a drag insensitive to
the sign of the potential (see Eq.~(\ref{e9})).

	From the result of Fig.~\ref{att_rep} it would be very interesting to redo
with an {\it attractive} perturbing potential the experiments which have been
done at M.I.T. with a repulsive potential \cite{Ram99,Ono00}. Although the
present computations are valid in a quasi one-dimensional regime (whereas the
experiments were done in truly three dimensional systems)
the present results leaves no doubt that the critical velocity for the onset of
dissipation should increase and that the energy transfer rate from the
obstacle to the fluid (i.e., $F_d\,v_\infty$) should drastically decrease in
case of an attractive obstacle \cite{rem}.

\

	Finally, we note that the present discussion sheds some light on the theory
of wave resistance. The wave resistance is the part of the drag experienced by
a body moving in a medium which is caused by excitation of waves in this
medium (typically surface waves in the case of boats). In a superfluid a $T=0$
({\it i.e.}, in the present work), this is the only source of drag (if one
broadens its definition in order to include non-linear effects such as vortex
or soliton formation). Recent experiments of moving spheres in silicone oil
\cite{Bur01} have shown that the wave resistance at Landau threshold has a
smooth behavior as a function of the velocity, contrarily to the expectation
of perturbation theory \cite{Rap96}. The same behavior was found here~:
non-linear effects smoothen out the unphysical step in drag predicted by
perturbation theory. This smooth behavior was already observed in the
experiments done at M.I.T. \cite{Ram99,Ono00} and in numerical simulations by
Frisch {\it et al.} \cite{Fri92} and Winiecki {\it et al.} \cite{Win99} when
moving an impenetrable sphere in a superfluid. Hence, Bose condensed systems
offer an interesting testing ground for ubiquitous non-linear hydrodynamical
effects, in a particularly simple theoretical framework (the Gross-Pitaevskii
equation).\\

\noindent {\large \bf Acknowledgments}

\bigskip \noindent It is a pleasure to thank P. Leboeuf and C. Schmit
for fruitful discussions.


\begin{thebibliography}{99}
\bibitem{attention} Note that we consider a quasi 1D system with infinite
coherence length, and discard from the present study quasi-condensates such as
observed by S. Dettmer {\it et al.}, Phys. Rev. Lett. {\bf 87}, 160406 (2001).
In fact, we can also describe quasi-condensates, the assumption made is that
the temperature is low enough so that the coherence length is large compared
to the size of the structure we describe (typically, a few times the
relaxation length).
\bibitem{dis} Even in the subsonic regime one finds stationary solutions with
a wake corresponding to dissipation. However, these solutions are presumably
unstable, whereas the solutions localized on the perturbation are found to be
numerically stable. Besides, when varying $v_\infty/c_\infty$ and the strength
of the potential, one finds that the dissipative sub-sonic solutions have a
smaller domain of existence than the superfluid ones: i.e., on the edge of the
domain of superfluidity, the localized dissipationless solutions are the only
existing ones (see Ref.~\cite{Leb01}).
\bibitem{trou} This is most clearly seen in the perturbative regime
$\kappa\sigma\gg 1$ and $V_0 \ll \kappa^{2}$, where (\ref{e8}) leads to:
$\delta A (x) = 2 \, \mbox{sgn}\, (v_\infty-c_\infty)\, \kappa^{-2}
\sqrt{n_\infty}\; V_\parallel(x)$.
\bibitem{rem} Moreover, the experiments can been done in the same quasi 1D
regime as the present theory, by taking as an obstacle a laser whose waist is
much larger than the transverse size of the condensate.
\end{thebibliography}
\end{document}